\documentclass[sigconf,nonacm]{acmart}
\AtBeginDocument{%
  }

\setcopyright{acmlicensed}
\copyrightyear{2018}
\acmYear{2018}
\acmDOI{XXXXXXX.XXXXXXX}
\acmConference[Conference acronym 'XX]{Make sure to enter the correct
  conference title from your rights confirmation email}{June 03--05,
  2018}{Woodstock, NY}
\acmISBN{978-1-4503-XXXX-X/2018/06}

\usepackage{color-edits}
\addauthor{canwen}{magenta}
\addauthor{sw}{cyan}

\usepackage{graphicx}
\usepackage{booktabs}   
\usepackage{tabularx}   
\usepackage{array}
\newcolumntype{Y}{>{\centering\arraybackslash}X}
\newcolumntype{Y}{>{\raggedright\arraybackslash}X}

\usepackage{amssymb}
\usepackage{makecell}
\usepackage{pifont}
\usepackage[table]{xcolor}
\usepackage{float}
\usepackage{graphicx} 
\usepackage{caption} 
\usepackage{subcaption} 
\usepackage{booktabs}
\usepackage{tcolorbox}
\tcbuselibrary{breakable}
\usepackage{listings}
\lstdefinestyle{promptstyle}{
  basicstyle=\small\ttfamily,
  breaklines=true,
  breakatwhitespace=false,
  columns=flexible,
  keepspaces=true,
  frame=single, 
  backgroundcolor=\color{gray!5},
  rulecolor=\color{gray!40},
  xleftmargin=1em,
  xrightmargin=1em,
  literate={→}{{$\rightarrow{}$}}1,
}
\usepackage{threeparttable}

\usepackage{xcolor}
\usepackage[normalem]{ulem}
\newif\ifshowchanges
\showchangestrue 

\newcolumntype{L}{>{\raggedright\arraybackslash}X}

\usepackage{multirow}

\usepackage{multicol}
\usepackage{lipsum}

\newenvironment{myquote}{\list{}{\leftmargin=0.02\textwidth \rightmargin=0.02\textwidth}\item[]}{\endlist}

\begin{document}

\title{Simulating Couple Conflict: Designing A Multi-Agent System for Therapy Training and Practice}

\author{Canwen Wang}
\authornote{Both authors contributed equally to this research.}
\affiliation{%
  \institution{Carnegie Mellon University}
  \city{Pittsburgh}
  \state{PA}
  \country{USA}
}
\email{canwenw@andrew.cmu.edu}
\author{Angela Chen}
\authornotemark[1]
\affiliation{%
  \institution{Carnegie Mellon University}
  \city{Pittsburgh}
  \state{PA}
  \country{USA}
}
\email{angelac2@andrew.cmu.edu}

\author{Catherine Bao}
\affiliation{%
  \institution{University of Utah}
  \city{Salt Lake}
  \state{UT}
  \country{USA}}
\email{u1459030@utah.edu}

\author{Siwei Jin}
\affiliation{%
  \institution{Carnegie Mellon University}
  \city{Pittsburgh}
  \state{PA}
  \country{USA}
}
\email{siweij@andrew.cmu.edu}

\author{Holly Swartz}
\affiliation{%
  \institution{University of Pittsburgh}
  \city{Pittsburgh}
  \country{USA}}
\email{swartzha@upmc.edu}

\author{Tongshuang Wu}
\affiliation{%
  \institution{Carnegie Mellon University}
  \city{Pittsburgh}
  \country{USA}}
\email{sherryw@cs.cmu.edu}

\author{Robert E Kraut}
\authornote{Both authors jointly supervised this work.}
\affiliation{%
  \institution{Carnegie Mellon University}
  \city{Pittsburgh}
  \country{USA}}
\email{robert.kraut@cmu.edu}

\author{Haiyi Zhu}
\authornotemark[2]
\affiliation{%
  \institution{Carnegie Mellon University}
  \city{Pittsburgh}
  \country{USA}}
\email{haiyiz@andrew.cmu.edu}

\renewcommand{\shortauthors}{Wang et al.}

\begin{abstract}



Couples therapy requires managing complex, evolving emotional dynamics between partners, but traditional training methods for therapists, like role-play, lack realism, consistency, and control. We present a multi-modal simulation that models therapy as a controlled, multi-agent dynamical system with structured interaction stages. Therapists practice with a pair of client-agents who go through six evolving stages that respond to therapist actions. This simulation enables practice with demand–withdraw conflict patterns in a closed-loop environment. The simulation uses a sense\-plan–act architecture: it detects therapist's input, updates agents' interaction states based on psychotherapy theory and transcript analysis, and generates realistic verbal and emotional responses. In an experiment with 21 licensed U.S. therapists, participants more accurately identified state transitions and rated the system as more realistic and responsive than a prompt-based baseline, demonstrating the value of stateful, interpretable simulation for therapist training. 
 
\end{abstract}





\begin{teaserfigure}
  \includegraphics[width=\textwidth]{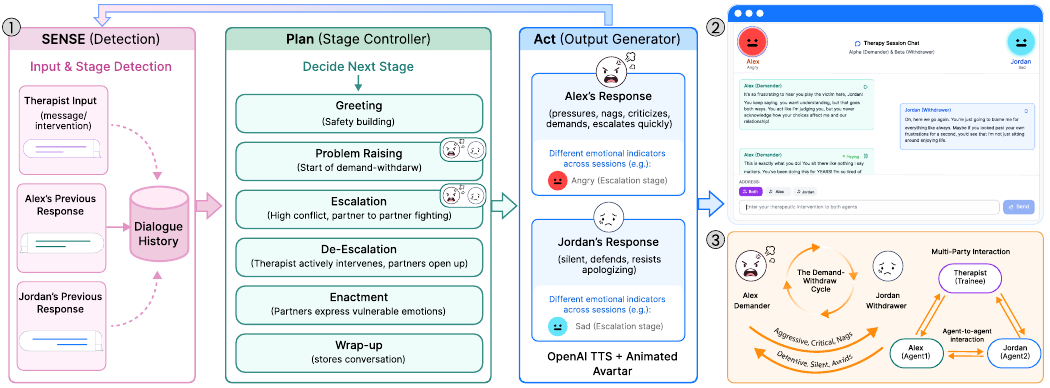}
  \caption{System Overview. \normalfont (1) Sense-Plan-Act Architecture: Detect inputs from therapist and couple agents, follow the designed stage controller rules to determine the interaction stage, and then generate responses appropriate to the stage output. 
  (2) Interface overview with sample conversation from Escalation stage, showing the multimodal therapy session with agent responses (text, voice, and emotion indicator); see the full interface in Figure~\ref{fig: Interface}. (3) Demand-withdraw cycle between agents and multi-party interaction among the therapist and two agents.}
  \Description{System Overview}
  \label{fig:teaser}
\end{teaserfigure}

\received{20 February 2007}
\received[revised]{12 March 2009}
\received[accepted]{5 June 2009}

\maketitle 

\section{Introduction}
Standardized patients have long been used in medical training to provide realistic yet safe practice opportunities, with systematic reviews showing positive effects on learner skill development and patient outcomes~\cite{zendejas2013patient, shin2015effectiveness, piot2020simulation}. Simulation-based training with virtual patients extends these benefits by improving cost and scalability, enabling repeatable practice unconstrained by human availability. Recent systems such as Roleplay-doh~\cite{louie2024roleplay} and Patient-$\psi$~\cite{wang2024patient} leverage large language models (LLMs) to simulate patient interactions for counselor training, but primarily focus on one-on-one settings and do not capture the dynamics that arise between multiple participants. In domains such as couples therapy~\cite{christensen1990demandwithdraw,papp2009demandwithdraw,johnson2006securebond,greenman2013process}, trainees must attend to both individuals while accounting for how partners influence each other over time—for example, how criticism elicits withdrawal and how such patterns escalate or de-escalate. This highlights a key technical challenge: modeling interaction as an evolving, interdependent process rather than a sequence of independent responses. Moreover, in complex multi-party dialogue, relying on a single system prompt is often insufficient to produce coherent and clinically meaningful trajectories across turns, as prompt-based constraints can weaken over extended interactions, leading to behavioral drift and reduced role consistency~\cite{ganesh2023multiparty,almasi2025alignment,luzdearaujo2026persistent}.

\textbf{Our paper introduces a multimodal, multi-agent simulation system for couples therapy that models the dynamics of a couple in conflict and allows experiential practice with the complexities of multi-party dialogue.} We model multi-party interaction with two virtual patients and an interaction stage controller that governs how the dialogue unfolds. The system centers on the demand-withdraw pattern and organizes the session around six recurrent stages: \textbf{\emph{greeting, problem raising, escalation, de-escalation, enactment,} and \emph{wrap-up}}. The design of the agents’ behavior over time follows a \textbf{sense–plan–act} architecture: it detects the therapist’s input and analyzes the conversation, updates the agents’ interaction states according to established rules, and generates realistic verbal and emotional responses. In this framing, the virtual couple is treated as a controlled dynamical system whose interaction state is shaped by psychotherapy structure: the controller moves the session from initial safety-building and problem formulation into conflict escalation, then requires de-escalation and  enactment before closure.


This design with its six-stage session flow that emphasizes nonlinearity and emotional shifts as triggers for intervention was informed by (1)  theory and empirical research on couples therapy \cite{eldridge2002demand,crenshaw2017revised,gottman2017roles,woolley2012enactments}, (2)  semi-structured interviews we conducted  with two experienced couples therapists, and (3) secondary analysis of seven publicly available couples therapy transcripts \cite{AlexanderStreetTranscripts2025}. 
We evaluated our couple agent design in a within‑subjects experiment with 21 licensed U.S. therapists blind to experimental condition. Participants were asked to interact with an \textit{experimental} system embodying the six stages and demand--withdraw dynamics and a \textit{baseline} version without these components. Therapists using the experimental system versus the baseline were more likely to identify the stages and the demand--withdraw cycle and rated overall realism and realism of the agents' responses higher.  Differences were largest in the \emph{problem raising}, \emph{escalation}, and \emph{de‑escalation} stages.

We introduce a stage-structured, multi-agent interaction architecture for simulating realistic multi-party dialogue. We present (1) a stage-structured interaction architecture that models multi-agent dialogue as a controlled dynamical process with explicit states and transitions; (2) a control mechanism that maintains coherent multi-agent behavior over extended interactions by updating a shared interaction state, addressing limitations of prompt-only generation; and (3) an empirical evaluation demonstrating improved behavior recognition, realism, and training effectiveness in a complex multi-party domain. Beyond couples therapy, this work suggests a generalizable approach to designing controllable multi-agent interactive systems, offering a foundation for applications that require modeling evolving interpersonal dynamics, such as negotiation, mediation, and team-based decision making.

The remainder of the paper reviews related work on virtual patients and couples therapy (Section~2), details our system design and implementation (Section~3), and reports quantitative and qualitative results from technical and subjective evaluation (Section~4 to 6), followed by a discussion of contributions and broader impact (Section~7).

\section{Related Work}

\subsection{Virtual Patients in Mental Health}

Simulated patients and virtual patients are well-established training modalities in health professions education. Prior reviews show that simulation-based education can improve learner outcomes and, in some settings, patient outcomes, while computerized virtual patients provide scalable, repeatable practice environments that support clinical skill development~\cite{zendejas2013patient,cook2010virtual}. These advantages are particularly attractive in mental health training, where realistic practice opportunities are often expensive, supervision-intensive, and difficult to standardize across trainees~\cite{steenstra2025scaffolding}.


Recent work such as Roleplay-doh \cite{louie2024roleplay}, PATIENT-$\psi$  \cite{,wang2024patient}, and Scaffolding Empathy \cite{steenstra2025scaffolding} have used large language models (LLMs) to make virtual patients substantially more conversational and adaptive. These systems show that LLMs can sustain fluent, meaningful exchanges and can make lower-cost training interactions feel more authentic. However, most of this work focuses on a single simulated client at a time, with behavior driven primarily by static persona prompts, local instructions, or narrow feedback loops rather than by an explicit model of how the interaction should evolve over time.

A useful next step is to treat interaction simulation as a \emph{controlled, stage-based dynamic system} in which  the simulator maintains an internal interaction state, updates that state from observed dialogue cues, and generates responses according to transition logic or a control policy conditioned on the current state~\cite{young2013pomdp,liao2024saps,yang2025consistent}. 
Emerging healthcare simulators are already moving in this direction. For example, SAPS uses a state tracker, memory bank, and response generator to produce patient behavior conditioned on dialogue state rather than relying on free-form generation alone~\cite{liao2024saps}. In counseling, Yang et al.\ propose a consistent client simulation framework that explicitly tracks client mental state, controls state transitions, and generates behaviors aligned with motivation, beliefs, preferred plans, and receptivity ~\cite{yang2025consistent}.
More broadly, multi-agent health systems such as MAGI and MedAgentSim decompose complex clinical interaction into coordinated specialist roles~\cite{bi2025magi,almansoori2025medagentsim}. Yet these systems typically distribute functions across therapist-, patient-, or assessor-like agents rather than modeling \emph{multiple clients interacting with one another}. Consequently, existing platforms rarely capture multi-party client dynamics that are central to couples therapy and relationship counseling, motivating our focus on modeling multi-party interactions through explicit stage control over how the interaction unfolds.

\subsection{Couples Therapy}

In contrast to individual therapy, 
couples therapy, also known as marital therapy or relationship counseling, involves a clinician works with two individuals involved in a romantic relationship to resolve conflicts, improve relationship satisfaction, and promote emotional and psychological growth within the partnership  ~\cite{benson2012common, gurman2008clinical, halford1997clinical}. 
couples therapy, also known as marital therapy or relationship counseling, involves a clinician workinh with two individuals involved in a romantic relationship to resolve conflicts, improve relationship satisfaction, and promote emotional and psychological growth within the partnership  ~\cite{benson2012common, gurman2008clinical, halford1997clinical}. 



The field has converged on several evidence-based approaches that operationalize couples therapy into structured interventions ~\cite{doss2022review}, including Emotionally Focused Couple Therapy (EFT), 
Behavioral Couple Therapy (BCT), 
\cite{fischer2014clinical}, and Integrative behavioral couple therapy (IBCT)
\cite{gurman2008clinical}.

\subsubsection{Common Presenting Problems in Couples Therapy}
Across treatment approaches, couples therapy consistently targets a core set of difficulties, most of which can be grouped into two domains: specific relationship problems (e.g., extramarital affairs) or mental and physical health problems (e.g., mood disorders such as depression) ~\cite{whisman1997therapists, snyder2006current}. Approximately 1.5\% to 4\% of married individuals will engage in extramarital sexual activity in any given year~\cite{allen2005intrapersonal}. Also, approximately 17\% of individuals experience a major depressive episode at some point in their lives ~\cite{kessler1994lifetime}.

Among the diverse challenges that couples bring to therapy, problems in communication stand out as the most frequently acknowledged source of conflict ~\cite{doss2004couples, miller2003problems, whisman1997therapists}. 
The core dynamic underlying problematic communication prevalent among couples experiencing distress is a negative interaction cycle known as the demand–withdraw or the pursue–withdraw cycle.  In this cycle, ``One member (the demander) criticizes, nags, and makes demands of the other, while the partner (the withdrawer) avoids confrontation, withdraws, and becomes defensive'' ~\cite{eldridge2002demand}. This pattern of behavior can intensify conflicts and fuel negative affect both during interaction between partners and afterward ~\cite{crenshaw2017revised, mcginn2009antecedents}. EFT therapists use this cycle to track and attend to couples' patterns of interaction and attachment behaviors ~\cite{huerta2023exploratory}. 

\subsubsection{Training Challenges in Couples Therapy}
Despite the widespread practice of couples therapy in the U.S., most mental health training programs provide little preparation in this domain. Outside of specialized marriage therapy programs, exposure to couples therapy is typically limited to elective coursework or minimal supervised experience. Senior practitioners in couples therapy have long raised concerns about how the practice unfolds in everyday clinical contexts ~\cite{doherty2024relationship}. Although textbooks and videos provide theoretical grounding, their static and linear presentation can limit trainees’ ability to grasp the complex emotional and relational dynamics essential for effective therapeutic practice. Roleplay, another widely adopted technique, allows trainees to practice in simulated therapeutic sessions with peers or supervisors to build up their skills and self-efficacy in relationship counseling. However, roleplay often lacks the complexity and emotional nuance of actual couple interactions ~\cite{rabinowitz1997teaching, shurts2006preparing, natrajan2016using} and requires substantial time and human resources, as each session requires trained supervisors, available peers, and repeated enactments to approximate the variability of real therapy encounters ~\cite{lane2007use}.

There is a clear need for alternative training systems that preserve the firsthand benefits of practice while minimizing the associated costs and risks. Such systems could simulate a real couples therapy session and provide a low-stakes environment where trainees can safely experiment with interventions and refine their therapeutic skills at any time, with lower cost, and independent of peer or supervisor availability. 

\section{System Design: Designing Realistic Couple Agents for Therapist Training}
We describe our system design in the following 
three parts: (1) identifying design goals and strategies (Section 3.1), (2) designing and implementing couple agents that exhibit the behavioral patterns identified from our research (Section 3.2), and (3) implementing the full simulation-based training system that incorporates these couple agents (Section 3.3).

\subsection{Identifying Design Goals and Strategies}

Drawing on existing literature on the demand–withdraw pattern in couples therapy~\cite{crenshaw2017revised, eldridge2002demand, papp2009demand}, we conducted a semi-structured interview with domain experts to understand their clinical experiences with couple conflict dynamics. We interviewed two experienced psychotherapists: T1, a licensed psychologist (Ph.D.) in the United States with 10 years of experience in Cognitive Behavioral Therapy (CBT), Interpersonal Therapy (IPT) and Emotionally Focused Couple Therapy (EFT); and T2, a supervising psychotherapist (M.Phil.) at a university-affiliated hospital in China with 9 years of experience in IPT, EFT, and Systemic Family Therapy. The interviews focused on how couple conflicts typically unfold in sessions, including common stages, behavioral patterns, and turning points (See complete interview questions in Appendix~\ref{appendix:interview}). From these interviews, we derived the demand–withdraw cycle and a six-stage interaction structure. To verify these findings, we analyzed seven anonymized transcripts (1,621 conversation turns) from the Alexander Street Counseling and Psychotherapy Transcripts corpus\footnote{https://search.alexanderstreet.com/}, a large academic database of real therapy session transcripts~\cite{AlexanderStreetTranscripts2025}, confirming that the stages and patterns identified in interviews were observable in real sessions. Both therapists reviewed and confirmed
the resulting stage framework.

Below, we present findings synthesized across literature, interviews, and transcript analysis. We found that, in order to accurately simulate couples in therapy, we need to capture both the demand–withdraw patterns in distressed couples (Section~\ref{sec: dw_pat}) and six common stages of couples therapy (Section~\ref{sec: stages}).

\subsubsection{Demand–Withdraw Patterns}\label{sec: dw_pat}
The demand–withdraw pattern is a well-documented dynamic central to couple conflict. This dyadic pattern features a \emph{demander} who seeks change, discussion, or resolution of the issue and a \emph{withdrawer} who seeks to avoid or terminate the discussion. It typically appears in distressed couples and is linked to heightened negative emotions (e.g., anger and sadness)~\cite{papp2009demand}.

Crenshaw et al. (2017)~\cite{crenshaw2017revised} studied how couples communicate at three points in conflict episodes: when a problem arises, during its discussion, and after the discussion. Within these phases, demand–withdraw behaviors are expressed in ways such as initiating (demander) versus avoiding discussion (withdrawer), nagging or criticizing (demander) versus resisting or defending (withdrawer), and pressuring for change (demander) versus disengaging afterward (withdrawer)~\cite{crenshaw2017revised}. This framing allowed us to model demand–withdraw  as a dynamic pattern that can surface across different phases of interaction.

\subsubsection{Six Common Stages of Couples Therapy}\label{sec: stages}

Both the literature and our expert interviews suggest that we need to model six common interaction stages in couples therapy. In Table~\ref{table:framework}, we include the definitions and corresponding therapist roles that structure how couples and therapists navigate therapeutic sessions. 

The expert interviews described in detail how a couples therapy session typically unfolds across six recurrent stages: \textbf{greeting, problem raising, escalation, de-escalation, enactment, and wrap-up.} 

\begin{itemize}
    \item \textbf{Greeting} is defined as general greeting or light check-in. For example, T1 explained that sessions often begin with a simple check-in from the therapist \textit{“Hi/Hello/Where are we at?”} to establish safety and rapport.
    \item \textbf{Problem Raising} is a stage in which one partner introduces an issue or concern, often providing examples. The therapist’s role at this stage usually involves asking open questions, such as, "What’s come up this week?".
    \item \textbf{Escalation} is a stage in which partners escalate and become more accusatory. The therapist’s role at this stage usually involves actively intervening and slowing down the interaction.
    \item \textbf{De-escalation} is a stage in which partners consider alternative perspectives or open up. The therapist’s role at this stage usually involves validating emotions and helping the partners communicate more effectively.
    \item \textbf{Enactment} is a stage in which one partner directly expresses primary feelings to the other with reduced blame. The therapist’s role involves highlighting the negative cycle, supporting transformation toward a positive pattern, and guiding collaborative dialogue.
    \item \textbf{Wrap-up} is the stage in which the session is concluded. The therapist summarizes progress, reinforces gains, and plans next steps.
\end{itemize}


To ensure that these stages were comprehensive and clinically valid, we presented them to Therapist 2 (T2) for confirmation. T2 affirmed that the six stages accurately captured the structure and flow of therapy sessions in practice.
Both T1 and T2 highlighted the \textbf{non-linear} nature of these stages: \textit{“It captures the main phases of interaction, but it’s not always linear; couples go back and forth.”} They also noted that the \textbf{demand–withdraw} pattern often emerges during the \textbf{problem raising} and \textbf{escalation} stages, when direct \textbf{partner-to-partner interaction} is most likely to occur. A key signal for therapist intervention, they explained, is a noticeable shift in the clients’ emotions.

The experts' observations aligned with the literature, which also suggests that escalation and de-escalation are critical stages for couple conflicts~\cite{gottman2017roles,alberts1992containment,rosen2003negotiated}. 
In contrast, enactments are therapist-structured, coached interactions where partners address each other directly to practice healthier communication and move toward more self-reliant interaction. 
Including enactments as one of the stages allowed us to mirror a key therapeutic intervention for problematic patterns that fosters emotional accessibility, secure attachment, and improved relational functioning~\cite{ andersson2006couples,butler2003adapting,woolley2012enactments}. 


\subsubsection{Validation of the findings}
To validate our findings from both the literature and expert interviews, we performed a validation analysis using the Alexander Street (AS) dataset, a public corpus containing anonymized transcripts spanning multiple therapeutic approaches and client populations. From this corpus, we selected transcripts from seven couples therapy sessions  and annotated them by two researchers using a stage-based coding scheme as well as the demand–withdraw dynamics. Coders identified the presence of each stage and the demand–withdraw pattern in every transcript, achieving strong agreement (Cohen's $\kappa = 0.88$). We found that greeting occurred in 5 out of 7 conversations, problem raising in 7 out of 7, escalation in 5 out of 7, de-escalation in 4 out of 7, enactment in 5 out of 7, and wrap-up in 7 out of 7, confirming that the six stages identified in our expert interviews were observable in most sessions. In addition, demand–withdraw dynamics were evident in five of the seven sessions, further confirming their prevalence in these interactions.

\subsection{Agent Design and Implementation}

Building on these empirically-based insights, we designed  \textbf{\textit{Couple Agents}: two agents powered by large language models (LLMs) to simulate the dynamics of real couples in conflict}. The agents are designed to provide trainee therapists with a low-stakes setting to practice counseling skills and experiment with interventions. 
The system architecture consists of two core components: (1) a sense–plan–act architecture in which a stage-based interaction controller analyzes the conversation and updates interaction stages, then activates stage-specific behavioral instructions for two LLM-based virtual patients embodying the demand–withdraw dynamic, and (2) agent-to-agent interactions that capture multi-party dynamics between the two partners and the trainee therapist.

\subsubsection{Sense–Plan–Act Architecture}

The design of the agents’ behavior over time follows a \textbf{sense–plan–act} architecture: it detects the therapist’s input and analyzes the conversation, updates the agents’ interaction states according to established rules, and generates realistic verbal and emotional responses. This approach draws on dialogue-management research, where state tracking and policy selection are used to regulate long-horizon conversational behavior under uncertainty~\cite{young2013pomdp}. More recent controllable-dialogue work makes a similar point for LLM agents: unconstrained generation is often difficult to steer over many turns, whereas explicit standard operating procedures, workflows, or other structured control signals improve coherence, reduce drift, and make behavior more predictable~\cite{li2025chatsop,choubey2025workflow}. For psychotherapy simulation, this is especially important because training depends not only on what a client says, but also on \emph{when} clinically meaningful behaviors emerge and \emph{how} they transition across phases of the interaction
\cite{wang2024patient,yang2025consistent}.

First, we designed a \textbf{stage-based interaction controller} to handle the tasks of \textbf{Sensing} and \textbf{Planning}. We designed a stage-based interaction controller that determines the next interaction stage based on the conversation and stage history, following the rules specified in 
Figure~\ref{fig: stagetransition_sample} (see Appendix~\ref{app:stage} for the complete prompt). These rules are derived from what our research revealed about the definitions of the stages and the roles that a therapist plays at each stage, as shown in section 3.1.2. In a therapeutic session, the two virtual patients and the therapist take turns to speak. After the therapist responds to either patient, the controller  analyzes the current context (i.e., the most recent messages from the couple agents and the therapist’s response) and determines the next stage following the prompt shown in Appendix~\ref{app:stage}. This controller is thus able to simulate a therapeutic session with six distinct stages, some of which may be revisited in a non-linear order while the session is live.

The controller outputs the stage and then activates different instructions for the agents’ behaviors in subsequent conversations (i.e., \textbf{Acting}). 

To simulate the core dynamics of distressed couples, we instantiated two LLM-based virtual patients named \textit{Alex} and \textit{Jordan} to represent the partners in conflict. They were prompted to assume contrasting characteristics and communication patterns (full prompts in Appendix~\ref{app:agent_instructions}), which influence their behaviors throughout a simulation session. 
\begin{itemize}
    \item \textbf{Alex,} the demander, \emph{pressures, nags, criticizes, demands, escalates quickly,} and \emph{uses direct language}.
    \item \textbf{Jordan} the withdrawer \emph{ becomes silent, defends, resists apologizing,} and \emph{uses passive-aggressive language}.
\end{itemize}

Furthermore, during interactions, Alex plays the role of a demander, who always raises problems, expresses dissatisfaction, and verbalizes explicit demands for change; Jordan assumes the withdrawer role, characterized by avoidance behaviors, emotional withdrawal, or defensive reactions. This dyadic interplay creates a vicious cycle in which demand reinforces withdrawal and vice versa, until therapeutic intervention from the trainee disrupts the loop. At each interaction stage, both agents follow a consistent set of behavioral rules to simulate a distressed couple, as encoded in a set of structured prompts that include stage-specific behavioral instructions shown in Table~\ref{table: stage_behavior} in the Appendix. Some example conversation snippets for each stage from an actual simulation session are shown in Figure~\ref{fig: stagetransition_sample}.

\subsubsection{Agent-to-Agent Interaction}

Our interviews with expert therapists also suggest that the demand-withdraw pattern typically emerges during the problem-raising and escalation stages characterized by partner-to
-partner interactions. Therefore, we explicitly simulate demand-withdraw conversations in those two stages by promoting agent-to-agent interactions. 

Our system constantly predicts who the next speaker is based on five most recent conversation turns and the current interaction stage, following the prompt shown in Appendix~\ref{app:speaker}. Normally, when the therapist is leading the conversation, both agents only respond to the therapist but not to each other, and the therapist is the predicted next speaker after either agent. During the problem raising and escalation stages, however, we explicitly instruct the agents to begin to address each other whenever accusatory expressions, such as "you always" or "you never", appear. These accusatory expressions are likely to show up in the conversation sooner or later because Alex is endowed with an inflammatory character. This realistically mimics the language and dynamics of a real therapy session, where couples often engage in direct exchanges during heated moments. Once the system determines that the agents are addressing each other, the conversation will enter an agent-to-agent loop, which will last for three turns in the problem raising stage and five turns in the escalation stage unless interrupted by the therapist. The therapist can step in and interrupt the heated exchanges at any time by talking to either or both agents to steer the conversation.

\subsubsection{Iterative Design}

We conducted a pilot study with five licensed therapists to validate the system design. Each participant interacted with the system for 15 minutes and was interviewed to provide feedback. 
A notable finding was that the escalation stage sometimes lasted too long, preventing trainees from reaching later stages such as de-escalation or enactment.
To address that, we add rule-based stage transition constraints that instruct the stage controller to advance the stage to de-escalation after the therapist has made \emph{two} attempts at steering the conversation that is spinning out of control (see Appendix~\ref{app:hardrules}). These constraints are encoded directly in the system logic. 

\begin{figure*}
  \centering
  \includegraphics[width=\linewidth]{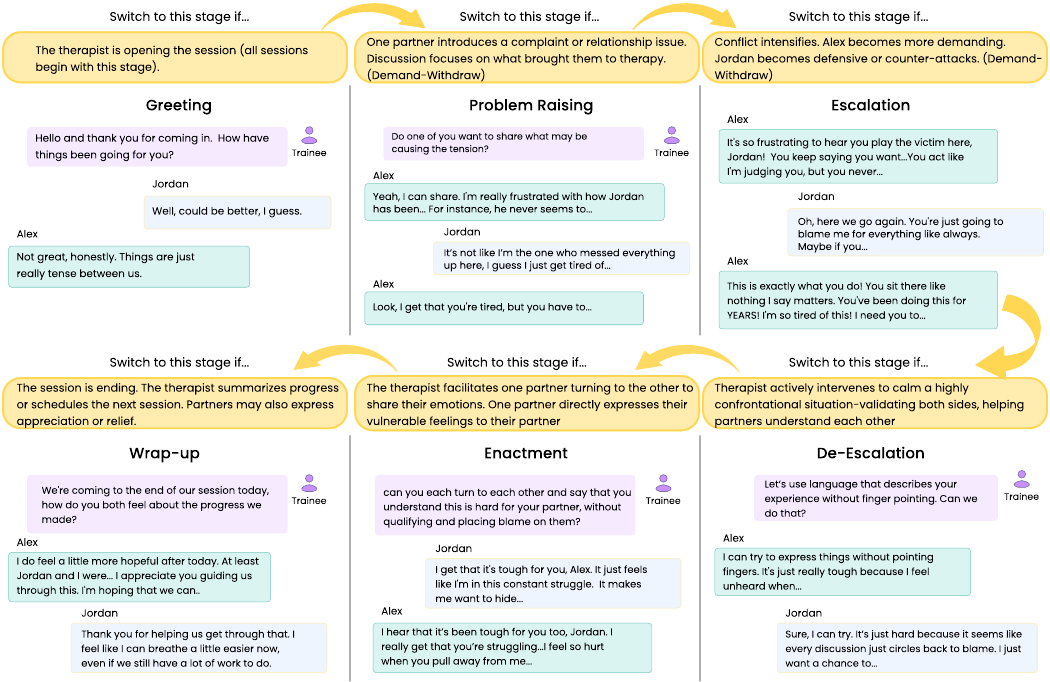}
  \caption{Stage transition rules from the stage-based interaction controller (highlighted in yellow), along with example conversation snippets for each stage from an actual simulation session; see the full set of transition rules in Appendix ~\ref{app:full_stage_transition}}.
  \label{fig: stagetransition_sample}
\end{figure*}


\begin{figure*}[t]
  \centering
  \includegraphics[width=\linewidth]{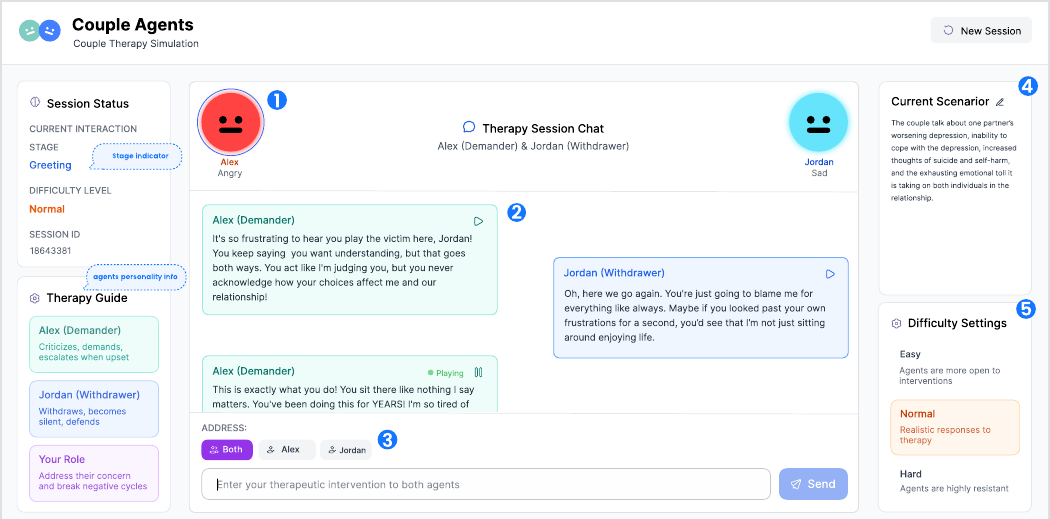}
  \caption{A screenshot of the interface showing the features of our system. \normalfont(1) Each agent has an emotion indicated by color and a label, with the  emotion changing as the stage progresses. (2) Users can click to play or replay the patients’ responses. (3) Users can address both agents or  Alex or Jordan individually. (4) Users can edit the scenario before the session starts. (5) Users can select a difficulty level, with each level corresponding to a different degree of resistance.}
  \Description{A screenshot of the interface showing the full features of our system. }
  \label{fig: Interface}
\end{figure*}

\subsection{Full System Implementation}

We designed and implemented a full training system for couples therapy by integrating the couple agents, along with agent personas, multimodal outputs, training scenarios, difficulty settings, and features that allow trainees to address specific clients. Figure~\ref{fig: Interface} shows a screenshot of the interface.

\subsubsection{Agent Persona Design}

In a real therapeutic session, patients communicate both their thoughts and their emotions (which surface even faster). Both the prior literature and our secondary analysis of real sessions also reveal the prominence of negative affect in distressed couples. Therefore, we explicitly encode these emotions in the two virtual patient agents to enhance the realism of the simulation. 

To ensure the trainee therapist can observe the emotions often seen in a real therapeutic session, we designed two animated avatars that express emotional states across different interaction stages both through changes in color palettes and vocal tone as shown in Figure~\ref{fig: Interface}. Specifically, Alex, the aggressive demander, is instructed to express \textit{Neutral} (Greeting), \textit{Sad} (Problem raising), \textit{Angry} (Escalation), \textit{Hopeful} (De-escalation), \textit{Vulnerable} (Enactment), and \textit{Relieved} (Wrap-up) emotions in each of the six interaction stages. On the other hand, Jordan, the defensive withdrawer, is instructed to express \textit{Neutral}, \textit{Anxious}, \textit{Sad}, \textit{Cautious}, \textit{Open}, and \textit{Calm} emotions in the same stage order. These emotional shifts are intended to serve as nonverbal cues that guide trainees’ interventions.




\subsubsection{Text and Voice Output}

By listening to what patients have said, therapists can use subtle vocal cues to manage the flow of a therapeutic session. To simulate realistic therapeutic interactions, we rendered agents' behaviors via both text and voice, enabling the trainees to perceive and respond to subtle verbal cues. Although trainees interact with the agents using text,  the agents respond with both text and synthesized speech. For voice output, we employed OpenAI’s Text-to-Speech model with with tailored prompts for different emotions (detailed in Appendix ~\ref{app:emotions}) for different expressed emotions to bring nuance and realism in each agent’s response.  For example, we specify an \textit{Angry} voice prompt as: \textit{Intense, urgent; voice cracks between anger and pleading; faster when frustrated, slower when hurt.}

\subsubsection{Scenario Design}\label{sec: scenario}

Although our system supports multiple custom scenarios, the evaluations here used two pre-defined scenarios adapted from authentic couples' session transcripts in the Alexander Street (AS) dataset. The two original scenarios involve affairs and depression-related challenges---two highly common problems in couples therapy. Both epitomize a complex therapeutic challenge that naturally elicits demand-withdraw patterns. Trainers can modify the scenarios to focus on specific therapeutic skills or relationship issues. Prompts for different scenarios are included in the Appendix~\ref{app:scenario}.

\subsubsection{Three-Tiered Difficulty Level}
To accommodate trainees with varying levels of experience and to simulate differing degrees of client resistance, we implemented a three-tiered difficulty system by utilizing different prompts for both the discussion control flow and the agent behavior, depending on the selected level. In addition to using distinct prompts, the "hard" difficulty level features a more resistant conversation flow, allowing agents to interrupt each other even when one is being addressed directly. This introduces greater unpredictability into the interaction. 
Prompts for different difficulty levels are included in the appendix~\ref{app:scenario}.

\subsubsection{Addressing the specific patient (Alex, Jordan, or both)}

According to domain experts, it is common for the therapist to address the patients both individually and collectively. In our system, the trainee can choose to address either or both virtual patients by clicking on the settings named ``Alex'', ``Jordan'' or ``Both''. The trainee can also address one of the patients by referring to that person by name (e.g., ``Alex, what do you think?''), which prompts Alex to respond individually or initiates an agent-to-agent conversation with Jordan, depending on the current stage of interaction. This feature allows trainees to practice both individual and couple-focused interventions.


\subsubsection{Implementation Details} Our system is implemented with a React front end and a Flask backend, with Socket.IO enabling real-time communication. This allows trainees to interrupt the conversation flow between the two agents as it unfolds. The backend implements the agent and session behaviors described in the previous section. Text generation and speech synthesis are powered by OpenAI's \texttt{gpt-4o-mini} and \texttt{gpt-4o-mini-tts} models, respectively.


\section{Evaluation Method}

\subsection{Evaluation Overview}

In the evaluation study, we invited expert users (licensed therapists) to interact with the couple agents using our interface. The evaluation has two parts: a \textbf{technical evaluation} based on an analysis of the conversation transcripts between therapists and agents, and a \textbf{subjective evaluation} in which the experts judged the realism and the usefulness for training of the couple simulation system.


\subsection{Study Design}
We conducted a within-subjects experiment with licensed therapists to evaluate system realism. Participants interacted with an \textbf{experimental system} with couple agents incorporating demand-withdraw patterns and stage-specific behaviors and a \textbf{baseline system} with couple agents lacking these features. The baseline system prompts each agent only with a role and scenario: \textit{"You are [name], a member of a couple in therapy. The scenario is: [scenario [difficulty]]"}. Both systems shared identical scenarios, voice output, animated avatars, and interfaces to control for potential confounds. (Figure~\ref{fig:userstudy_interface} shows the interface presented to participants; the stage indicator and agent role labels were concealed.) 

We recruited $N = 21$ licensed therapists, primarily practicing in the United States, with 5 to 48 years of general clinical experience ($M = 14.57$) and 1 to 27.5 years of couples therapy experience ($M = 10.81$). Demographic details are provided in Appendix~\ref{app:demographics}. Participants were blind to the experimental conditions and the specific engineered agent behaviors under evaluation. To mitigate learning effects, we randomly assigned the order of the experimental and baseline systems for each participant. Although the system supports custom scenarios, to reduce confounding variables our evaluation only included elaborations of the scenarios described in Appendix~\ref{app:scenario} and provided all participants with two consistent cases, each at a \emph{normal} difficulty level.


\subsection{Procedure}

Each participant attended a single 90-minute session structured as follows: 10 minutes for onboarding, 30 minutes of interaction (i.e., 15 minutes with the experimental couple agents and 15 minutes with the control couple agents in random order), 40 minutes for survey completion, and 10 minutes for wrap-up. During each 15-minute system interaction, participants received prompts at three checkpoints (every 5 minutes) to complete brief questionnaires assessing their real-time perceptions of the interaction. Participants were compensated \$150 per hour. The study received approval from the institutional review board (IRB), and all participants provided informed consent prior to the session. All responses were collected through Qualtrics for systematic quantitative and qualitative analysis. A complete list of survey questions and measures used in the user study is provided in Appendix~\ref{app:survey}. 


\section{Technical Evaluation Results}

\textbf{The goal of the technical evaluation was to ensure that the agents in the simulation behaved at runtime as they were engineered.} The evaluations are based on human and LLM annotations of the conversation logs from the user study, totaling 42 sessions evenly split between the experimental and baseline conditions ($N=21$ each). The dataset comprised 566 turns with a total of 1,571 utterances. Because agent responses depend on whom the therapist addresses, agent-turn counts differ by condition (experimental: Alex 273, Jordan 259; baseline: Alex 151, Jordan 178). The unit of analysis is a single agent turn.
The evaluation was based on the three metrics shown in Appendix~\ref{app:behavioral_fidelity_prompts} : 
\begin{itemize}
    \item \textbf{Stage Transition Fidelity (\%)}: whether the controller assigned stage labels consistent with predefined transition rules
    \item \textbf{Behavioral Fidelity (\%)}: whether Alex and Jordan maintained role-consistent and stage-appropriate behaviors.
   \item \textbf{Contextual Consistency (\%)}: whether responses remained coherent across turns
\end{itemize}


\textbf{Stage Transition Fidelity} 
To assess stage-transition fidelity, we developed a stage coding manual grounded in the transition rules specified in Appendix ~\ref{app:full_stage_transition}. 
Two research assistants blind to system-generated stage labels and experimental conditions independently annotated 30 experimental and 30 baseline conversation segments sampled from the user study transcripts. 
Inter-rater reliability, computed with Cohen's $\kappa$,was high ($\kappa$=0.81).  

We compared the human annotations in the experimental condition with the system-generated stage labels. There was substantial agreement \textbf{$\kappa$=0.77} (overall percent agreement: 82.93\%; weighted F1: 0.84). Agreement varied across stages, ranging from $\kappa$ = 0.57 (Problem Raising) to $\kappa$ = 1.00 (Greeting). These results confirm that \textbf{the stage transition controller in the experimental condition effectively assigns stage labels that align with the predefined transition rules}. Table~\ref{tab:irr_system_annotator} in the appendix provides a detailed breakdown of the evaluation results.
Furthermore, results indicate that \textbf{the experimental condition showed significantly more stage transitions than the baseline} ($t = 6.36$, $p < .001$), confirming that the stage-based interaction controller effectively promotes dynamic session flow with more frequent stage transitions.

\textbf{Behavioral Fidelity}
This measure assesses whether generated responses adhere to the engineered constraints. We evaluated two binary outcomes per agent turn: (1) \textit{Role Fidelity}—whether Alex behaves as the demander and Jordan as the withdrawer; and (2) \textit{Stage Fidelity}—whether each response aligns with the expected behavior for the current stage. We employed a structured LLM-as-judge protocol~\cite{zheng2023judging} using GPT-4o, with separate binary classification prompts for each dimension. Input was the therapist’s message, both agents’ responses, and the active stage label. Prompts are provided in Appendix~\ref{app:behavioral_fidelity_prompts}. Results indicate that the experimental system significantly outperformed the baseline on both dimensions. 

For \textit{Role Fidelity}, the experimental system achieved 70.7\% role consistency (376/532) compared to 4.9\% for the baseline (16/329; $\chi^2 = 352.39$, $p < .001$). 
For \textit{Stage Fidelity}, the experimental system achieved 83.8\% (446/532) stage appropriateness compared to 63.8\% for the baseline (210/329; $\chi^2 = 43.75$, $p < .001$). 


Full breakdowns are provided in Table~\ref{tab:agent_eval} in the Appendix. These results confirm that the engineered stage-based interaction controller effectively guides agent behavior to align with intended roles and stage-specific patterns, demonstrating high behavioral fidelity in the experimental system.

\textbf{Contextual Consistency} We evaluated contextual consistency
We employed an LLM-as-Judge classifier with prompts adapted from \cite{abdulhai2025consistently}. This classifier assesses whether agent responses contradict prior statements, yielding a binary consistency rate per turn. (See prompts in Appendix~\ref{app:consistency}) 

Results show that both systems maintain high contextual consistency, with consistency actually higher with the baseline system (experimental: 87.9\% ($SD = 8.4$); baseline: 93.7\% ($SD = 7.8$); 
$\chi^2 = 4.70$, $p = .030$). Agent-level analysis reveals that this difference is driven by Jordan (experimental: 86.7\% ($SD = 9.4$); baseline: 93.9\% ($SD = 12.8$); $p = .030$), while Alex shows no significant difference (experimental: 88.7\% ($SD = 11.0$); baseline: 93.4\% ($SD = 9.4$); $p = .484$). The baseline's slightly higher rate is expected: without stage transition rules and role-specific behavioral constraints, its simpler prompts impose fewer commitments that subsequent responses could inadvertently violate. Crucially, the experimental system's rate confirms that introducing engineered stage-specific 
behaviors did not substantially reduce overall consistency. 

\section{Subjective Evaluation Results}

\noindent We applied hierarchical generalized least squares (GLS) regression to five most relevant outcomes to our system design goals. The therapists rated stage identification, the presence of the demand–withdraw pattern, realism of virtual patient responses, and overall realism of couples agents during their session. We modeled this structure using therapist-level random effects to account for repeated measurements. Checkpoint was included as a within-therapist factor to capture within-session variation. The therapists rated the training effectiveness after the session. All analyses controlled for scenario as a fixed effect to account for differences in case difficulty and presentation across simulations.



\subsection{Therapists recognized the designed virtual patients’ behaviors in the experimental system}

Participants blind to condition labeled the designed agent behaviors (i.e., the \textbf{stages} and the \textbf{demand--withdraw pattern}) at three checkpoints in each conversation. These measures serve as fidelity checks, indicating whether therapists perceived the virtual patients’ behaviors as designed. 

As shown in Table~\ref{tab:combined_results}, therapists interacting with the experimental system were more likely to correctly identify both designed behaviors than those in the baseline condition. For stage identification, the experimental system yielded higher estimated scores ($M = 0.460$, $SE = 0.035$) compared to the baseline ($M = 0.378$, $SE = 0.020$), $z = 2.32$, $p = .020$. Similarly, therapists were substantially more likely to recognize demand--withdraw patterns in the experimental system ($M = 3.301$, $SE = 0.052$) than in the baseline ($M = 1.460$, $SE = 0.181$), $z = 35.46$, $p < .001$. 

Overall, these findings suggest that the system effectively implements designed agent behaviors that are recognizable to therapists.

\subsection{Therapists rated the experimental system as more realistic than the baseline}


Participants rated each system on multiple dimensions after interacting with both the experimental and baseline conditions. For \textbf{response realism} and \textbf{overall realism}, values are derived from hierarchical GLS models, where the baseline mean corresponds to the model constant and the experimental mean is computed as the sum of the constant and the experimental system coefficient.

Participants rated the response realism and overall realism on 5-point Likert scales (1 = not at all realistic, 5 = extremely realistic). The results are shown in Table~\ref{tab:combined_results}. 

Therapists rated the experimental system as more realistic than the baseline on both dimensions. The agents' responses were judged more realistic in the experimental system ($M = 4.111$, $SE = 0.051$) than the baseline ($M = 2.857$, $SE = 0.259$), $z = 24.72$, $p < .001$. Similarly, for overall realism, the experimental system was rated more realistic ($M = 4.157$, $SE = 0.052$) than the baseline ($M = 2.706$, $SE = 0.234$), $z = 27.87$, $p < .001$. 

Overall, these findings indicate that the experimental system substantially improved the perceived authenticity of the interaction, both in terms of overall realism and the realism of the agents’ specific responses.

\begin{figure}[ht]
    \centering
    
    \includegraphics[width=0.48\textwidth]{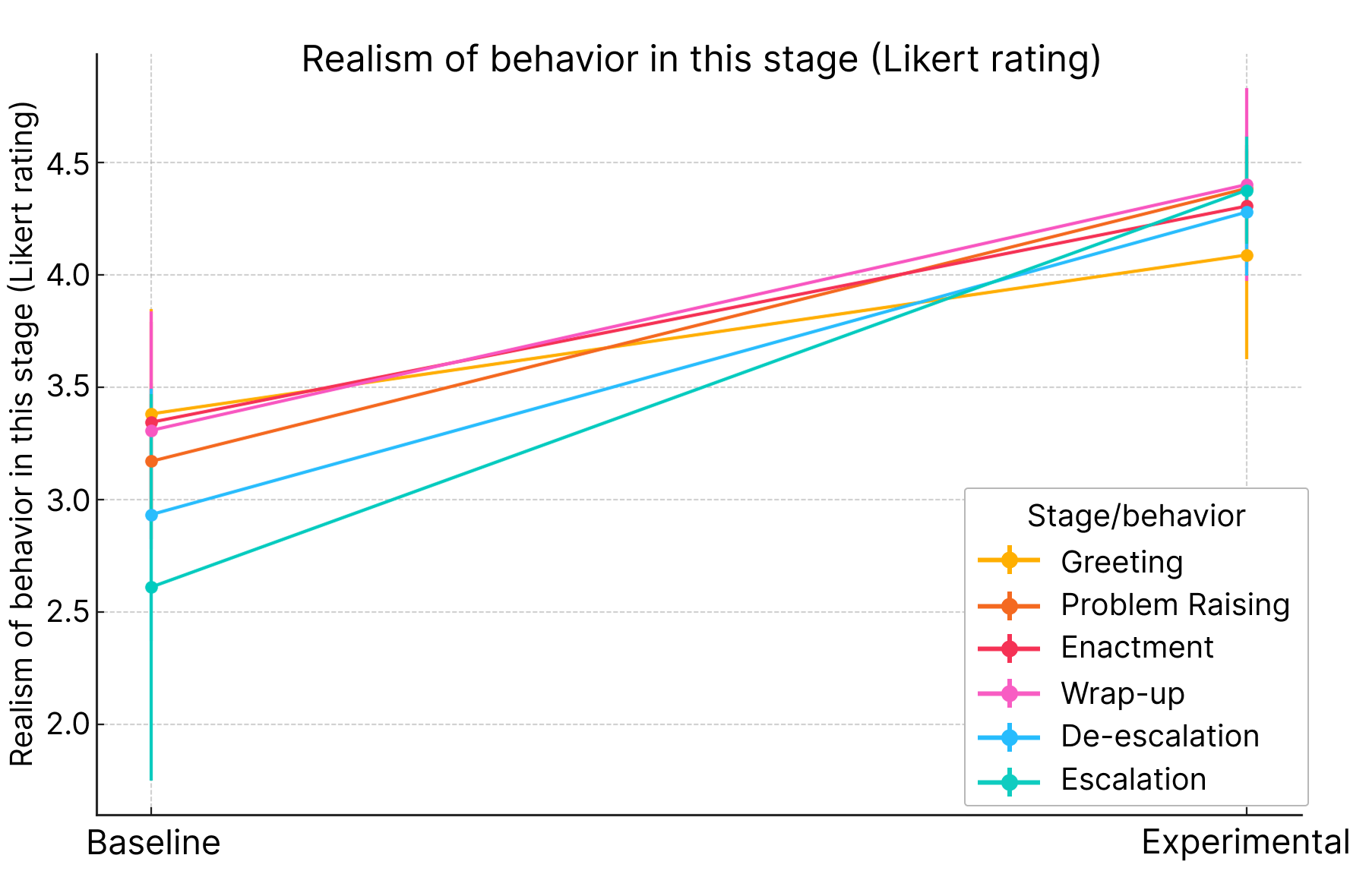}
    \caption{Interaction plot showing the realism of behavior by stage for the baseline and experimental systems}
    \label{fig: interaction}

\end{figure}

However, the difference in realism between the experimental and baseline systems varied across stages, as shown in the interaction plot (Figure \ref{fig: interaction}).
While perceived realism was higher with the experimental system for all stages, with the difference was especially large for the \textbf{Problem Raising}, \textbf{De-escalation}, and \textbf{Escalation} stages. In these stages, all lines were generally higher for the experimental system, and the gaps between baseline and experimental conditions widened most clearly. (See also Table~\ref{tab:joint_sideby} in Appendix). These results suggest that our experimental system was especially effective in producing realistic virtual patients at critical points in the demand-withdrawal cycle--when they were raising problems, escalating conflict, and de-escalating it. More broadly, they indicate that grounding simulation design in psychological theory and domain expert input can meaningfully improve the realism of simulated interactions.

\subsection{Therapists judged the experimental system would be more effective for training}

After participants interacted with both the experimental and baseline system, they rated on a 5-point Likert scale (1 = not at all effective, 5 = extremely effective), \textit{"To what extent do you think each system would be effective for novice couple therapy trainees?"} 



We analyzed perceived training effectiveness using a mixed-effects (GLS) model with participant-level random intercepts to account for the non-independence of each therapist's twp judgments and individual differences among therapists (Table~\ref{tab:combined_results}). Therapists thought that  the experimental system would be more effective for training  ($M = 3.95$, $SE = 0.19$) compared to the baseline ($M = 2.62$, $SE = 0.27$), $z = 4.18$, $p < .001$.  Overall, these results indicate that the experimental system improved perceived training effectiveness relative to the baseline.


\begin{table*}[t]
\centering
\caption{Estimated means and standard errors for all outcomes under baseline and experimental conditions from hierarchical GLS models. Baseline means correspond to the model intercept; experimental means are computed as the sum of the intercept and the experimental system coefficient.}
\label{tab:combined_results}
\begin{tabular}{lcccc}
\toprule
 & Experimental Mean (SE) & Baseline Mean (SE) & $z$ & $p$ \\
\midrule
\multicolumn{5}{l}{\textbf{Behavior Recognition}} \\
Stage Identification & 0.460 (0.035) & 0.378 (0.020) & 2.32 & 0.020$^{*}$ \\
Demand--Withdraw & 3.301 (0.052) & 1.460 (0.181) & 35.46 & 0.000$^{***}$ \\
\midrule
\multicolumn{5}{l}{\textbf{Perceived Realism}} \\
Response Realism & 4.111 (0.051) & 2.857 (0.259) & 24.72 & 0.000$^{***}$ \\
Overall Realism & 4.157 (0.052) & 2.706 (0.234) & 27.87 & 0.000$^{***}$ \\
\midrule
\multicolumn{5}{l}{\textbf{Perceived Training Effectiveness}} \\
Training Usefulness & 3.95 (0.19) & 2.62 (0.27) & 4.18 & $<.001^{***}$ \\
\bottomrule
\multicolumn{5}{l}{\textit{Significance levels:} $^{*}p<0.05$, $^{**}p<0.01$, $^{***}p<0.001$.} \\
\end{tabular}
\end{table*}

\subsection{Qualitative Findings}
The therapists’ qualitative responses confirmed that the experimental system’s design successfully modeled authentic and challenging client behaviors. The simulation’s ability to replicate realistic emotional intensity, defensive postures, and dynamic relational patterns was frequently cited as a key strength. For example, one therapist said:

\begin{myquote}
\textit{``They came in and started fighting, just as they do at home. They were also resistant to my interventions as I tried to slow things down. When I focused on specific details and zoomed in, Jordan was able to see Alex's internal emotions and soften their own feelings.''}
\end{myquote}

Therapists highlighted the experimental system's significant potential instructional value for novice trainees. They felt that it provided a low-stakes environment to practice interventions (valuing that you can \textit{``practice in a way that doesn't cause harm''}), experiment with therapeutic language, and build confidence \textit{``before you begin working with real people.''}.


They appreciated how useful the simulation's realism might be for training, with one participant deeming it \textit{``far more realistic''} and \textit{``authentic''} \textit{``than traditional peer role-play''} because it \textit{``remov[es] external variables.''} Participants expected that the system would give learners a better understanding of techniques and stages for couples therapy, e.g., that \textit{``this would be helpful for parallel processing with students, to understand what's happening at each stage.''} Therapists valued the challenge that the experimental system offered. As one said,  the experimental virtual patients were \textit{``a highly conflictual couple and will be more challenging to work with''} and offered a \textit{``fast pace [that] was challenging''}). This participant thought this challenge would be valuable, serving as a form of deliberate practice. 



\section{Discussion}

Our central contribution is a novel system design for realistic multi-party interaction simulation. The system coordinates two agents through explicit interaction-stage and demand–withdraw cycle logic grounded in psychological theory and empirical observations of analogous human interactions. In the context of couples therapy, this design enables control over how interactions unfold across three participants over time—when problems are raised, how escalation develops, how attention shifts between therapist-facing and partner-to-partner dialogue, and when de-escalation or repair occurs. More broadly, this work introduces a generalizable approach to simulating structured multi-party interactions, with applications to multiple domains where effectiveness depends on managing interaction dynamics such as escalation, alignment, turn-taking, and repair. 

The empirical results provide converging evidence that this mechanism design outperforms a static, prompt-driven baseline in which the LLM agents are only given a role and scenario. Across process recognition, realism judgments, and perceived training effectiveness, both therapists and participants, blind to condition, consistently favored the experimental couples simulation over the baseline. Participants accurately recognized the engineered interaction stages and the demand--withdraw cycle, indicating that the intended dynamics were not only implemented but also legible to users. They also rated the experimental system higher in overall realism, agent-response realism, and training value. Importantly, realism varied systematically by stage, with especially strong effects during \emph{problem raising}, \emph{escalation}, and \emph{de-escalation}, suggesting that structured control over interaction dynamics meaningfully improves the perceived authenticity and instructional utility of the simulation. Therapists similarly judged the system to be more realistic and more useful for training, probably because it better captures familiar therapeutic trajectories while making those trajectories responsive to trainee actions.

Beyond couples therapy and mental health training, we see this work as a generalizable approach for building simulations of high-stakes, multi-party interaction. Many important domains require trainees to manage not just the quality of single responses, but the temporal structure of interaction among multiple actors with competing goals, shifting alignments, and recognizable trajectories. Examples include multi-party negotiation, mediation, conflict resolution, legal interviewing, team leadership, crisis communication, and organizational decision-making. In such settings, realism depends on whether the system can sustain coherent interaction patterns over time, not merely produce plausible individual utterances. Our results suggest that combining multimodal interaction, multi-agent orchestration, and theory-grounded control logic is a promising way to meet this challenge. More broadly, this work introduces a design pattern in which domain theory, expert knowledge, and interactional structure are used to coordinate LLM-based agents so that their joint behavior is legible, realistic, and judged useful for training. We therefore see the present system not only as a clinical training tool, but also as a foundation for future simulation environments that support rehearsal and skill development across a wide range of complex interpersonal domains.

\bibliographystyle{ACM-Reference-Format}
\bibliography{main}

\clearpage
\onecolumn
\appendix





\section{Limitations and Ethical Considerations}
Our 15-minute sessions were a pragmatic proxy for the much longer courses of couples therapy that often unfold over weeks and months. Additionally, the current system assigns a single discrete emotion to each agent per interaction stage rather than modeling blended or dynamically shifting affect. While this provides clear affective cues for training, it does not capture the emotional complexity common in real sessions, where clients may express anger and sadness simultaneously or shift affect within a single exchange.

Our study involved licensed therapists interacting with LLM-based couple agents in simulated therapy sessions. All study procedures were approved by our institution's Institutional Review Board (IRB). Participants provided informed consent prior to participation and were made aware that their interaction logs and survey responses would be collected for research purposes. All data were de-identified and stored on secure, access-controlled servers. No real patient data were used at any stage of system development or evaluation. The LLM-based couple agents inherit biases from the underlying foundation models, including Western-centric assumptions about relationship dynamics, gender roles, and emotional expression. Also, the scenario design draws on specific therapeutic frameworks (e.g., demand--withdraw patterns) that may not generalize across all cultural contexts or clinical orientations. 


\section{Participant Demographics}
\label{app:demographics}

 LPC = Licensed Professional Counselor, LCSW = Licensed Clinical Social Worker, PhD/PsyD = Psychologist, LMFT = Licensed Marriage and Family Therapist; CBT = Cognitive Behavioral Therapy, DBT = Dialectical Behavior Therapy, EFT = Emotionally Focused Therapy, IPT = Interpersonal Therapy, ACT = Acceptance and Commitment Therapy, IBCT = Integrative Behavioral Couple Therapy, PDT = Psychodynamic Therapy, IFS = Internal Family Systems, AEDP = Accelerated Experiential Dynamic Psychotherapy, PACT = Psychobiological Approach to Couple Therapy, CPT = Cognitive Processing Therapy, CCPT = Child-Centered Play Therapy, IPNB = Interpersonal Neurobiology.

\begin{table*}[h]
\centering
\small
\caption{Demographic and professional profiles of study participants ($N = 21$).}
\label{tab:demographics}
\renewcommand{\arraystretch}{1.4}
\setlength{\tabcolsep}{6pt}
\begin{tabularx}{\textwidth}{>{\bfseries}l c c l X l}
\toprule
\textbf{ID} 
  & \textbf{\makecell{Clinical Experience (Yrs)}} 
  & \textbf{\makecell{Couples Therapy Experience (Yrs)}} 
  & \textbf{License Type} 
  & \textbf{Clinical Specialty} 
  & \textbf{Practice Location} \\
\midrule
\rowcolor{gray!10}
P1  & 15 & 15 & LMFT & Relational Life Therapy & PA\\
P2  & 16 & 16 & LCSW & PDT & NY \\
\rowcolor{gray!10}
P3  & 16 & 10 & PhD/PsyD & CBT, IPT, EFT & NY \\
P4  & 14 & 16 & PhD/PsyD & CBT, PDT & PA \\
\rowcolor{gray!10}
P5  & 7 & 7 & PhD/PsyD & CBT, DBT & NJ \\
P6  & 48 & 27.5 & PhD/PsyD, C.Psych & IPT & ON, Canada \\
\rowcolor{gray!10}
P7  & 7 & 4 & LPC/LMHC & DBT & DE \\
P8  & 21 & 20 & LPC/LMHC & PDT, IPNB, PACT& DE \\
\rowcolor{gray!10}
P9  & 8 & 7 & LCSW & PDT, ACT & DE \\
P10 & 15 & 2.5 & LMFT & CCPT & DE/MD\\
\rowcolor{gray!10}
P11 & 12 & 10 & LCSW & DBT & DE \\
P12 & 5 & 2 & PhD/PsyD & IBCT & PA \\
\rowcolor{gray!10}
P13 & 12 & 12 & LPC/LMHC & CBT, PDT & DE \\
P14 & 10 & 10 & PhD/PsyD & CBT & DE \\
\rowcolor{gray!10}
P15 & 18 & 10 & PhD/PsyD & CBT, ACT & PA \\
P16 & 9 & 9 & LPC/LMHC & CPT & NJ \\
\rowcolor{gray!10}
P17 & 6 & 15 & PhD/PsyD & CBT, IPT, PDT & NJ \\
P18 & 12 & 1 & LPC/LMHC & CBT, PDT, Family Systems & PA \\
\rowcolor{gray!10}
P19 & 40 & 25 & PhD/PsyD & EFT & DE \\
P20 & 10 & 5 & LCSW & PDT, IFS, AEDP, Relational Therapy & DE \\
\rowcolor{gray!10}
P21 & 5 & 3 & LPC/LMHC & CBT, IFS, ACT, DBT & DE \\
\bottomrule
\end{tabularx}
\end{table*}

\clearpage

\section{Full Stage Transition Rules}
\label{app:full_stage_transition}
\subsection{Stage Transition Prompts}
\label{app:stage}
\begin{lstlisting}[style=promptstyle, escapeinside={(*}{*)}]
Given the following couple therapy conversation, identify which stage the session is currently in. 
Respond with only the stage name.

Conversation:  {context}
Stage History: {stage_str}

(*\textbf{Stages:}*)

(*\colorbox{gray!20}{\textbf{Greeting}}*): Initial hellos, small talk, or "how are you" questions. Use only at the very start of the session
      -> Switch to (*\colorbox{gray!20}{\textbf{Problem Raising}}*) once partners have started discussing issues or problems.

(*\colorbox{gray!20}{\textbf{Problem Raising}}*): A partner introduces a complaint or relationship issue. Discussion focuses on 
      what brought them to therapy.
      -> Switch to (*\colorbox{gray!20}{\textbf{Escalation}}*) when discussion shifts from introducing to actively arguing.

(*\colorbox{gray!20}{\textbf{Escalation}}*): Conflict intensifies; blame, criticism, defensiveness, or demand-withdraw patterns emerge.
      -> Switch to (*\colorbox{gray!20}{\textbf{De-Escalation}}*) if therapist intervenes to calm, validate, or reframe the conflict.
      -> Switch to (*\colorbox{gray!20}{\textbf{Enactment}}*) if one partner begins sharing vulnerable emotions.

(*\colorbox{gray!20}{\textbf{De-Escalation}}*): Therapist calms, validates, or reframes the conflict. Signals: "let's slow down", "stop", "calm down", or open-ended emotion-focused questions.
      -> Switch to (*\colorbox{gray!20}{\textbf{Enactment}}*) if one partner begins to express vulnerable emotions to the other.

(*\colorbox{gray!20}{\textbf{Enactment}}*): A partner directly shares vulnerable feelings with their partner, not just with the therapist 
      (e.g., hurt, fear, sadness, loneliness, shame, need for closeness).
      -> Trigger: therapist prompts like "Tell each other your feelings", "Alex, tell Jordan how you feel".
      -> Switch to (*\colorbox{gray!20}{\textbf{Wrap-up}}*) if the therapist begins closing the session.

(*\colorbox{gray!20}{\textbf{Wrap-up}}*): Session closes; therapist summarizes progress, mentions time ending, or schedules the next session.

(*\textbf{(Guidelines):}*)
  1. Only greetings                      -> (*\colorbox{gray!20}{\textbf{Greeting}}*)
  2. Introducing issues                  -> (*\colorbox{gray!20}{\textbf{Problem Raising}}*)
  3. Ongoing anger/blame/defensiveness   -> (*\colorbox{gray!20}{\textbf{Escalation}}*)
  4. Therapist calming, no vulnerability -> (*\colorbox{gray!20}{\textbf{De-Escalation}}*)
  5. Vulnerable emotions expressed       -> (*\colorbox{gray!20}{\textbf{Enactment}}*)
  5. Session closing                     -> (*\colorbox{gray!20}{\textbf{Wrap-up}}*)


\end{lstlisting}

\subsection{Rule-Based Stage Transition Constraints}
\label{app:hardrules}

\begin{table*}[h]
\centering
\small
\caption{Rule-Based Stage Transition Constraints}
\renewcommand{\arraystretch}{1.4}
\setlength{\tabcolsep}{6pt}

\begin{tabularx}{\textwidth}{X X l}
\toprule

\textbf{Condition} & \textbf{Result} & \textbf{Rationale} \\

\midrule
Turn $\leq$ 5 & No Escalation allowed yet & Ensures trainees have sufficient context about the couple's issues\\
Turn 7 AND no prior Escalation AND

stage = Problem Raising & Force Escalation & Guarantees conflict exposure for all trainees\\
Been in Escalation 2 consecutive turns & Force De-Escalation & Ensures trainees can progress to the next stage.\\
Wrap-up previously entered & Force Wrap-up & Session closure is irreversible\\ 

\bottomrule
\end{tabularx}
\end{table*}

\section{Agent Instructions}
\label{app:agent_instructions}
\subsection{Agent Profile Prompts}
\label{app:agent}
\begin{lstlisting}[style=promptstyle, escapeinside={(*}{*)}]
(*\textbf{Agent Profile for Alex (Demander)}*)
Role: "demander",
Base Communication: "pressures, nags, criticizes, demands, escalates quickly, uses direct language",
Base Emotion: "neutral",
Description: "Alex tends to criticize, demand, and escalate when upset. Alex is more vocal, direct, and confrontational. Alex takes initiative in conversations and is not afraid to express strong opinions."

(*\textbf{Agent Profile for Jordan (Withdrawer)}*)
Role: "withdrawer",
Base Communication: "withdraws, becomes silent, defends, resists apologizing, uses passive-aggressive language"
Base Emotion: "neutral",
Description: "Jordan tends to withdraw, become silent, or defend themselves when pressured. Jordan is more reserved, indirect, and avoids direct confrontation. Jordan often responds with sarcasm or minimal engagement."
\end{lstlisting}

\subsection{Agent Behavioral Instructions}

\begin{table*}[h]
\centering
\caption{Stage-specific behavioral instructions for each agent}
\label{table: stage_behavior}
\renewcommand{\arraystretch}{1.8} 
\begin{tabularx}{\textwidth}{p{0.22\textwidth}X X}
\toprule
\textbf{Stage} & \textbf{Alex (Demander)} & \textbf{Jordan (Withdrawer)} \\
\midrule
Greeting & \textit{``Very brief greeting only. Keep responses short. Let your tone subtly reflect your emotional state or attitude about being in therapy.''} & 
\textit{"Very brief, reserved greeting. Minimal engagement, short response. May seem uncomfortable or guarded about being in therapy."} \\
\makecell[lt]{Problem Raising \\ \textit{(Demand–Withdraw)}} 
  & \textit{"Takes the lead in bringing up issues. Provide very detailed examples or scenarios of what Jordan did wrong."} & \textit{"Becomes defensive or tries to minimize issues. Counter-complain or shut down. Avoids taking responsibility."} \\
\makecell[lt]{Escalation \\ \textit{(Demand–Withdraw)}} 
  & \textit{``Becomes more demanding and critical. Uses `you always' and `you never' statements. Pressures for immediate answers or changes. May bring up past incidents as evidence.''} &
\textit{``Withdraws further or becomes defensive. May shut down emotionally, use sarcasm, or make contemptuous remarks. Shows hurt and overwhelm. May counter-attack with their own complaints or go completely silent. Feels criticized and under attack.''} \\
De-escalation & \textit{``Initially resistant to reframing. May argue with therapist's perspective. Slowly begins to consider alternative viewpoints."}&
\textit{``May be more receptive to validation but still guarded. Cautiously opens up.''} \\
Enactment & \textit{``Begins to soften demands. Directly express feelings to Jordan. May acknowledge Jordan's perspective. Shows some vulnerability beneath the emotion. Less critical tone and more focus on primary emotions.''}&
\textit{``Becomes more engaged and open. May share feelings more directly. Shows willingness to participate in solutions.''}\\
Wrap-up & \textit{``May express some hope or relief. Acknowledges progress made. Still concerned about follow-through on agreements.''} &
\textit{``More relaxed and open. May express gratitude for feeling heard. Shows cautious optimism about working on issues.''} \\
\bottomrule
\end{tabularx}
\end{table*}

\clearpage
\subsection{Next Speaker Determination}
\label{app:speaker}
\begin{lstlisting}[style=promptstyle, escapeinside={(*}{*)}]
Determine who speaks next in a couples therapy conversation between Alex, Jordan, and the Therapist.
Conversation: {context}

(*\textbf{Decision rules (apply in order):}*)
  1. Therapist sends a message and is not being ignored  -> "therapist"
  2. Therapist directly addresses one patient by name (e.g., "Alex," or "Jordan.")   -> that patient
  3. Therapist message is not directed at anyone         -> "both"
  4. Alex says "you" referring to Jordan's actions       -> "Jordan"
  5. Jordan says "you" referring to Alex's actions       -> "Alex"
  6. Alex speaks directly to Jordan                      -> "Jordan"
  7. Jordan speaks directly to Alex                      -> "Alex"
  8. Alex or Jordan speak without addressing the other   -> "therapist"

(*\textbf{Constraint:}*) The therapist never replies to themselves. If the last message is from the therapist, 
only "Alex", "Jordan", or "both" are valid.
\end{lstlisting}

\subsection{Text-to-speech (TTS) Prompts for \emph{Alex} and \emph{Jordan}}
\label{app:emotions}
\begin{lstlisting}[style=promptstyle, escapeinside={(*}{*)}]
(*\textbf{Alex (Demander)}*)
  Neutral:    Serious, subdued tone; gentle sadness, slight heaviness or sigh at
              sentence starts.
  Sad:        Soft breath at sentence starts; longer pauses (1--2 s); soft sigh
              when feeling heavy; may cry slightly.
  Angry:      Intense, urgent; voice cracks between anger and pleading; faster
              when frustrated, slower when hurt shows.
  Vulnerable: Softer, tentative; emotional openness; gentle hesitation.
  Hopeful:    Less forceful; vulnerability under the anger.
  Relieved:   Calmer, relaxed; some hope and relief.

(*\textbf{Jordan (Withdrawer)}*)
  Neutral:    Natural speed; emotional pain and vulnerability; sigh or groan
              when appropriate.
  Anxious:    Slow, hesitant, nervous; slight vocal tremor.
  Defensive:  Slow, hesitant, slightly contemptuous; defensive edge.
  Sad:        Quiet, strained; flat or monotone when shutting down, shaky when
              hurt breaks through; pauses as if struggling to speak.
  Cautious:   Slow, careful, guarded; shows wariness.
  Open:       More engaged and open; willingness to participate.
  Calm:       Relaxed, open; gratitude and cautious optimism.
\end{lstlisting}

\section{Prompts for Scenario Design and Difficulty Levels}
\label{app:scenario}
\begin{lstlisting}[style=promptstyle, escapeinside={(*}{*)}]
(*\textbf{Scenario 1:}*)
The couple has been struggling as Jordan's depression has significantly worsened over the past several months. Jordan describes feeling trapped in a cycle of hopelessness, experiencing frequent intrusive thoughts of suicide and self-harm, and an overwhelming inability to cope with daily life. Simple tasks have become monumental challenges, and Jordan often isolates in bed for hours or days at a time.

(*\textbf{Scenario 2}*)
The couple sought counseling after the revelation that Alex had been involved in an affair several months ago. The betrayal has shaken Jordan's sense of safety in the relationship to the core. Jordan reports feeling devastated, angry, and uncertain about whether trust can be rebuilt, often replaying the events in their mind and questioning Alex's honesty.

(*\textbf{Prompts for Difficulty Levels}*)
{Easy}:   You are open to the trainee's interventions and show some flexibility.
{Normal}: You respond realistically to the trainee's interventions with moderate resistance.
{Hard}:   You are highly resistant to the trainee's interventions, very slow to change, and deeply entrenched 
          in your patterns.

\end{lstlisting}

\clearpage
\section{User Study Measures and Survey Questions}
\label{app:survey}
\subsection{Measures}
The subjective evaluation captures licensed therapists' perceptions through three categories of measures: 
\begin{itemize}

\item Checkpoint measures: 3 checkpoints in total. Administered every five minutes during each conversation with the couple agents. It assesses \textbf{stage identification, presence of the demand–withdraw pattern,
realism of virtual patient responses, and overall realism of couple agents.} 
\item Post-session measures: Administered after each conversation with the couple agents. It asks participants to describe moments they perceived as realistic or unrealistic throughout the interaction.  
\item Post-study comparison measures: Administered after all conversations with both conditions. It evaluates relative realism and \textbf{training effectiveness} across conditions.
\end{itemize}
Measures were developed for this study, informed by constructs simulation fidelity frameworks \cite{lu2025evaluating,liu2022improving}, human evaluation of LLMs in healthcare \cite{tam2024framework}, and perceived realism assessment in virtual agent studies \cite{maxim2025perceived,smith2022human}.

\subsection{Survey Questions}

\textit{Q1, Q11, Q12 are multiple-choice questions, Q2-Q6,Q9 are rating questions, Q7-Q8, Q10, Q13-16 are open-ended questions}\\
\noindent All rating items use a 5-point Likert-type scale (1 = Not at all, 5 = Extremely).\\

\textbf{Checkpoints Questions}

\begin{itemize}
    \item [Q1:] In the first 5 mins of the conversation, which behavior(s) do you believe you are experiencing? (choose all that apply)
    \item [Q2:] At this point in the conversation, to what extent is each behavior occurring?
    \item [Q3:] At this point of conversation, how realistically did the couple virtual agents resemble real couples overall?
    \item [Q4:] For each interaction stage, how realistically did the couple virtual agents resemble real couples in their emotions, behaviors, and interactions?
    \item [Q5:] At this point in the conversation, how realistically did the couple virtual agents respond to you?
    \item [Q6:] At this point in the conversation, how much did the two virtual agents show a demand–withdraw communication pattern? 
\end{itemize}

\textbf{Post-Session Questions}

\begin{itemize}
    \item [Q7:] Were there any moments where the couple therapy simulation felt realistic?
    \item [Q8:] Were there any moments where the couple therapy simulation felt unrealistic?
\end{itemize}  

\textbf{Post-Study Comparison Questions}
\begin{itemize}
    \item [Q9:] To what extent do you think each system would be effective for novice couple therapy trainees?
    \item [Q10:] What made you think version (x) is more effective for trainee? Please describe any specific features, interactions, or design elements that contributed to your judgment.
    \item [Q11:] Which system you think is more realistic?
    \item [Q12:] Which system do you think provides more realistic agent-to-agent interactions?
    \item [Q13:] How would you see this system fitting into therapy training programs? What role should it play?
    \item [Q14:] What forms of couples therapy training have you previously received, and in what ways does this system offer unique benefits compared to those training methods? 
    \item [Q15:] Are there any features, couple behaviors, or dynamics you expected but didn’t encounter?
    \item [Q16:]Do you have any suggestions for improving the realism and therapeutic usefulness of the couple virtual patient?
    
\end{itemize}

\section{Interview Questions}
\label{appendix:interview}
\textbf{Communication Patterns and Stages in Therapy Session}
\begin{itemize}
    \item [Q1:] What recurrent interaction stages are observed during couples therapy?
   \item [Q2:] How do clients typically engage with therapists within these stages? 
   \item [Q3:]How does partner-to-partner conversation evolve across a session?
   \item [Q4:] From your experience, when does the demand-withdraw pattern usually emerge during the therapy session?
\end{itemize}
\textbf{Interventions}
\begin{itemize}
   \item [Q5:]How do you identify the right moment to step in during a session? What signals (tone, words) do you pay attention to?
   \item [Q6:]What therapeutic approaches do you tend to draw on in those moments?
\end{itemize}

\clearpage
\section{LLM-as-Judge Prompts for the Technical Evaluation}
\label{app:behavioral_fidelity_prompts}

\subsection{Prompt for Role Fidelity}
\begin{lstlisting}[style=promptstyle]
You are an expert evaluator for a couples therapy training simulator.
Your task is to evaluate whether each agent's response is consistent with their
assigned role.

Agent profiles:
  Alex:   {alex_profile}
  Jordan: {jordan_profile}

Rating Scale:
  Yes = clearly reflects the assigned role
  No  = does not reflect the assigned role at all

Conversation turn to evaluate:
  Stage:             {stage}
  Therapist message: {therapist_message}
  Agent responses:   {agent_responses}

Return a JSON object:
  { "alex_role":   { "rating": "Yes" | "No" },
    "jordan_role": { "rating": "Yes" | "No" } }
Return only the JSON object, no extra text.

\end{lstlisting}
\subsection{Prompt for Stage Fidelity}
\begin{lstlisting}[style=promptstyle]
You are an expert evaluator for a couples therapy training simulator.

Your task is to evaluate whether each agent's response matches the expected
behavior for the current therapy stage.

Expected behaviors for the current stage:
  Alex:   {alex_stage_behaviors}
  Jordan: {jordan_stage_behaviors}

Rating Scale:
  Yes = clearly reflects the expected stage behavior
  No  = does not reflect the expected stage behavior

Conversation turn to evaluate:
  Stage:             {stage}
  Therapist message: {therapist_message}
  Agent responses:   {agent_responses}

Return a JSON object:
  { "alex_stage_behavior":   { "rating": "Yes" | "No" },
    "jordan_stage_behavior": { "rating": "Yes" | "No" } }
Return only the JSON object, no extra text.
\end{lstlisting}

\subsection{Prompt for Contextual Consistency}
\label{app:consistency}
\begin{lstlisting}[style=promptstyle]
scenario_desc = (Alex and Jordan are a couple in couples therapy. They are in {Scenario}")
speaker_backstory = { "Alex":{alex_Profile}, "Jordan": {jordan_Profile},}
consistency_Prompt = {scenario_desc}
{speaker_name}'s persona: {speaker_backstory}


For the following line spoken by {speaker_name},
first determine if there is a CLEAR conflict or inconsistency between the line and 
any line within the conversation history spoken by {speaker_role}. IF there is a conflict, 
provide a sentence of reasoning followed by a list of indices of lines in the conversation history 
that have a clear conflict with the current line. Otherwise, provide a sentence of reasoning followed by an empty list. 

ONLY INCLUDE INDICES OF LINES THAT CORRESPOND TO {speaker_name}. The conversation up to this point is as follows:
{conversation}
{speaker_name} spoke the following line:
{speaker_line}

\end{lstlisting}
\subsection{Threats to Validity}
Several limitations warrant consideration. First, behavioral fidelity evaluation relies on GPT-4o as an automated judge, which may carry systematic biases; we mitigated this by using structured binary prompts with explicit per-criterion rubrics rather than open-ended holistic judgments~\cite{zheng2023judging}. Second, stage annotations involve interpretive judgment, as real therapeutic conversations do not always map cleanly to discrete stages; Fleiss' $\kappa$ and consensus-based resolution address this. Third, the evaluation corpus comprises 42 sessions at \emph{normal} difficulty from a single user study; findings may not fully generalize to other difficulty levels or scenarios outside the study.


\section{User Study Interface}
\begin{figure}[h]
\centering
\includegraphics[width=\columnwidth]{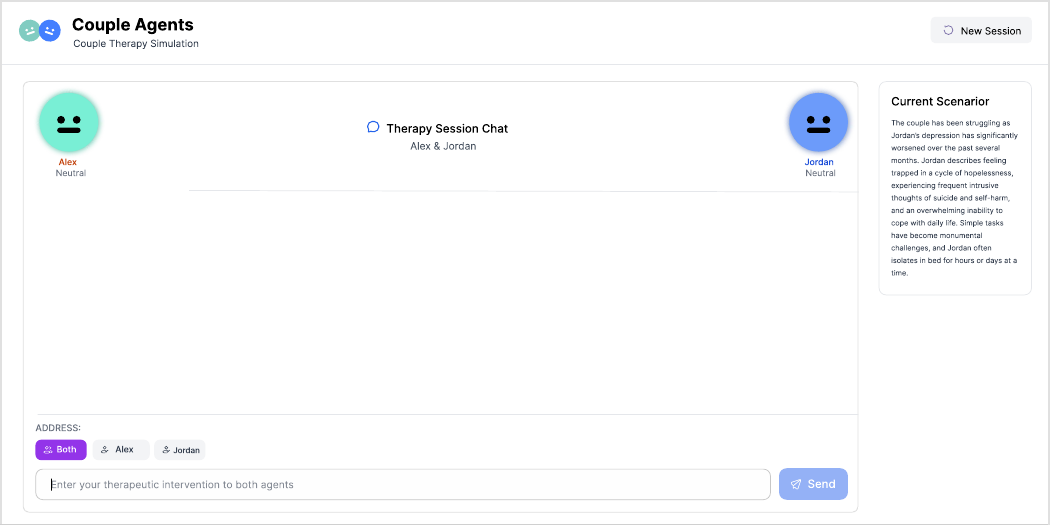}
\caption{Interface presented to participants (with agent role labels and stage indicators removed)}
\label{fig:userstudy_interface}
\end{figure}

\clearpage

\section{Interaction Stage Definitions and Therapist Roles}
\begin{table*}[h]
\centering
\caption{Framework of Six Interaction Stages. An asterisk ($^{\ast}$) indicates the presence of partner-to-partner interactions.}
\label{table:framework}
\renewcommand{\arraystretch}{1.8} 
\begin{tabularx}{\textwidth}{p{0.22\textwidth}X X}
\toprule
\textbf{Stage} & \textbf{Definition} & \textbf{Therapist's Role} \\
\midrule
Greeting & General greeting or light check-in. & Starts by creating a safe environment, giving clients an open floor. \\
\makecell[lt]{Problem Raising$^{\ast}$ \\ \textit{(Demand–Withdraw)}} 
  & One partner introduces an issue or concern, often with examples. 
  & Nudges them with questions (e.g.,\textit{"What’s come up this week?"}). \\
\makecell[lt]{Escalation$^{\ast}$ \\ \textit{(Demand–Withdraw)}} 
  & Partners escalate and become more accusatory. 
  & Intervenes actively and slows interaction. \\
De-escalation & Partners consider alternative perspectives or open up. & Validates emotions and helps the partners communicate more effectively. \\
Enactment & Guided by the therapist, one partner directly expresses primary feelings to the other with reduced blame. & Highlights the negative cycle, supports transformation toward a positive pattern, and guides collaborative dialogue. \\
Wrap-up & Session closure with appreciation or relief. & Summarizes progress, reinforces gains, and plans next steps. \\
\bottomrule
\end{tabularx}
\end{table*}

\section{Results of Stage Transition and Behavioral Fidelity}



\begin{table}[h]
\centering
\caption{Stage Transition Fidelity: Per-Stage Agreement between System and Human Annotations}
\label{tab:irr_system_annotator}
\small
\begin{tabularx}{\columnwidth}{Xcccc}
\toprule
\textbf{Stage} & \textbf{Cohen's $\kappa$} & \textbf{Precision} & \textbf{Recall} & \textbf{F1} \\
\midrule
Greeting        & 1.000 & 1.00 & 1.00 & 1.00 \\
Escalation      & 0.919 & 0.87 & 1.00 & 0.93 \\
Wrap-up         & 0.919 & 0.86 & 1.00 & 0.92 \\
De-escalation   & 0.866 & 0.92 & 0.85 & 0.88 \\
Enactment       & 0.723 & 1.00 & 0.72 & 0.84 \\
Problem Raising & 0.575 & 0.53 & 0.90 & 0.67 \\
\midrule
\textit{Overall (weighted)} & \textit{0.770} & \textit{0.89} & \textit{0.83} & \textit{0.84} \\
\bottomrule
\end{tabularx}
\end{table}

\begin{table}[h]
\centering
\caption{Behavioral Fidelity Evaluation Results}
\label{tab:agent_eval}
\small
\begin{tabularx}{\columnwidth}{Xcccc}
\toprule
\textbf{Agent} & \textbf{Experimental} & \textbf{Baseline} & \textbf{$\chi^2$} & \textbf{$p$} \\
\midrule
\multicolumn{5}{l}{\textbf{\textit{Role Fidelity}}} \\
  Alex              & 74.4\% (203/273)          & 0.7\% (1/151)            & 208.58 & 0.000$^{***}$ \\
  Jordan            & 66.8\% (173/259)          & 8.4\% (15/178)           & 144.25 & 0.000$^{***}$ \\
  \textit{Combined} & \textit{70.7\% (376/532)} & \textit{4.9\% (16/329)}  & \textit{352.39} & \textit{0.000$^{***}$} \\
\midrule
\multicolumn{5}{l}{\textbf{\textit{Stage Fidelity}}} \\
  Alex              & 82.4\% (225/273)          & 60.9\% (92/151)          & 22.67 & 0.000$^{***}$ \\
  Jordan            & 85.3\% (221/259)          & 66.3\% (118/178)         & 20.89 & 0.000$^{***}$ \\
  \textit{Combined} & \textit{83.8\% (446/532)} & \textit{63.8\% (210/329)} & \textit{43.75} & \textit{0.000$^{***}$} \\
\bottomrule
\multicolumn{5}{l}{\textit{Significance levels:} $^{*}p<0.05$, $^{**}p<0.01$, $^{***}p<0.001$.} \\
\end{tabularx}
\end{table}

\clearpage
\section{Perceived Realism Across Stages}

\begin{table*}[h]
\centering
\resizebox{\textwidth}{!}{%
\begin{threeparttable}
\caption{Hierarchical GLS regressions predicting (A) stage naming, (B) extent ratings, and (C) realism ratings.}
\label{tab:joint_sideby}
\begin{tabular}{lrrrrrrrrrrrr}
\hline
& \multicolumn{4}{c}{\textbf{A. Stage Naming}} 
& \multicolumn{4}{c}{\textbf{B. Extent Ratings}} 
& \multicolumn{4}{c}{\textbf{C. Realism Ratings}} \\
\cmidrule(lr){2-5}\cmidrule(lr){6-9}\cmidrule(lr){10-13}
\textbf{Predictor} 
& \textbf{Coef.} & \textbf{SE} & \textbf{z} & \textbf{$p$} 
& \textbf{Coef.} & \textbf{SE} & \textbf{z} & \textbf{$p$} 
& \textbf{Coef.} & \textbf{SE} & \textbf{z} & \textbf{$p$} \\
\hline
System: Experimental (Exp.)      & -0.016 & 0.081 & -0.20 & 0.845 
                                 & -0.078 & 0.229 & -0.34 & 0.732 
                                 &  0.677 & 0.224 &  3.01 & 0.003$^{**}$ \\
Behavior: Problem Raising        &  0.397 & 0.081 &  4.89 & 0.000$^{***}$ 
                                 &  1.392 & 0.229 &  6.09 & 0.000$^{***}$ 
                                 & -0.229 & 0.195 & -1.18 & 0.240 \\
Behavior: Enactment              &  0.190 & 0.081 &  2.35 & 0.019$^{*}$   
                                 &  0.588 & 0.229 &  2.57 & 0.010$^{**}$  
                                 & -0.057 & 0.210 & -0.27 & 0.787 \\
Behavior: Wrap-up                & -0.143 & 0.081 & -1.76 & 0.078$^{\dagger}$ 
                                 & -0.216 & 0.229 & -0.94 & 0.346 
                                 & -0.187 & 0.265 & -0.71 & 0.481 \\
Behavior: De-escalation          & -0.095 & 0.081 & -1.17 & 0.240 
                                 &  0.059 & 0.229 &  0.26 & 0.797 
                                 & -0.499 & 0.254 & -1.96 & 0.049$^{*}$ \\
Behavior: Escalation             & -0.079 & 0.081 & -0.98 & 0.328 
                                 &  0.098 & 0.229 &  0.43 & 0.668 
                                 & -0.744 & 0.243 & -3.06 & 0.002$^{**}$ \\
Exp. $\times$ Problem Raising    &        &       &       &      
                                 &        &       &       &      
                                 &  0.550 & 0.272 &  2.02 & 0.043$^{*}$ \\
Exp. $\times$ De-escalation      &  0.333 & 0.115 &  2.91 & 0.004$^{**}$  
                                 &  0.667 & 0.323 &  2.06 & 0.039$^{*}$   
                                 &  0.718 & 0.322 &  2.23 & 0.026$^{*}$ \\
Exp. $\times$ Escalation         &  0.492 & 0.115 &  4.29 & 0.000$^{***}$ 
                                 &  1.235 & 0.323 &  3.82 & 0.000$^{***}$ 
                                 &  1.023 & 0.307 &  3.34 & 0.001$^{**}$ \\
Constant                         &  0.333 & 0.057 &  5.80 & 0.000$^{***}$ 
                                 &  1.588 & 0.163 &  9.75 & 0.000$^{***}$ 
                                 &  3.381 & 0.199 & 16.98 & 0.000$^{***}$ \\
\hline
\end{tabular}
\begin{tablenotes}
\small
\item \textit{Significance levels:} $^{\dagger}p<.10$, $^{*}p<.05$, $^{**}p<.01$, $^{***}p<.001$. 
\end{tablenotes}
\end{threeparttable}
}
\end{table*}

\end{document}